\documentclass[10pt,aps,twocolumn,nofootinbib]{revtex4-1}
\usepackage{amsfonts}
\usepackage{mathrsfs}
\usepackage{amsmath}
\usepackage{amssymb}
\usepackage[normalem]{ulem}
\usepackage{extarrows}
\usepackage{slashed}
\usepackage{isodateo}
\usepackage{graphicx} 
\usepackage{xcolor}
\usepackage[bookmarksnumbered=true,bookmarksopen=true]{hyperref}
 \hypersetup{colorlinks,%
             linkcolor=[rgb]{0,0.3,0.6}, %
             citecolor=[rgb]{0,0.3,0.6}, %
             urlcolor=[rgb]{0,0.3,0.6}}%
\renewcommand{\rm}{\mathrm} 
\begin{document}

\title{Illuminating sequential freeze-in dark matter with dark photon signal at the CERN SHiP experiment}

\author{Xinyue Yin}
\altaffiliation{yinxy@stu.cqu.edu.cn}
\affiliation{Department of Physics, Chongqing University, Chongqing 401331, China}

\author{Sibo Zheng}
\altaffiliation{sibozheng.zju@gmail.com}
\affiliation{Department of Physics, Chongqing University, Chongqing 401331, China}

\begin{abstract}
Single-field freeze-in dark matter barely leaves observable footprints in  dark matter direct detection, collider or fixed-target experiments,
which can be altered in the two-field context. 
In this work, we consider sequential freeze-in dark matter through signals of dark photon mediator with a mass range of $m_{A'}\sim 10^{-2}-10$ GeV  covered by the proposed SHiP experiment.
We show that the dark charge is fixed to be $e'\sim 1.3\times 10^{-12}$ and the mixing parameter is restricted to $10^{-11}\leq \epsilon< 10^{-8}-10^{-7.5}$, 
as a result of the out-of-equilibrium condition of dark photon and the observed relic abundance of dark matter.
Within this $\epsilon$ region, the 5(15)-year data of proton bremsstrahlung process for the dark photon, 
assuming vector meson (dipole) dominance, excludes $\epsilon\geq 10^{-8.5} (10^{-7.9})$ at 90\% confidence level,
implying only a narrow region of $\epsilon$ close to $\sim 10^{-11}$ left for alternative tests.
\end{abstract}
\maketitle

\section{Introduction}
Dark matter (DM) is a key ingredient of standard cosmology.
Its particle nature is still unknown.
If DM couples to the Standard Model (SM) sector with a coupling strength being near unity, 
it is able to keep thermal equilibrium with the SM thermal bath, 
and obtains the observed relic abundance \cite{Planck:2018vyg} via so-called freeze-out mechanism. 
Different from the freeze-out mechanism, 
freeze-in mechanism \cite{Hall:2009bx} or a mixture of them \cite{Chu:2011be}, 
specialized for the DM coupling strength being tiny,
has gained a renewed interest recently due to null experimental results on the freeze-out DM scenarios. 

Well-known single-field  examples of freeze-in DM include millicharged particle \cite{Holdom:1985ag,Feldman:2007wj},
sterile neutrino \cite{Asaka:2005cn,Becker:2018rve,Datta:2021elq}, scalar \cite{McDonald:2001vt}, 
and axion-like particle \cite{Xu:2024cof}. 
Such freeze-in DM barely leaves observable footprints in collider or DM detection experiments, 
except that only few of them are in the reaches of cosmic ray experiments such as Ly$\alpha$ or 21-cm signals within limited parameter regions \cite{Xu:2024uas}.

Beyond the single-field situation, two-field freeze-in DM model can be in the scopes of collider, fixed target/beam dump or DM detection experiments,
as illustrated by earlier studies \cite{Hessler:2016kwm,Ghosh:2017vhe,Calibbi:2018fqf,Belanger:2018sti,Yin:2024sle} on a fermion or scalar mediator communicating between the DM and SM sector.
The main reason is that coupling strength of the mediator to the SM sector is not necessarily as small as the DM coupling to the mediator.

Inspired by the aforementioned studies, 
we consider freeze-in DM through a massive dark photon mediator; 
see  e.g., \cite{Fabbrichesi:2020wbt} for a recent review.
Notably, this two-field freeze-in DM model differs from a millicharged DM via massless dark photon \cite{Kovetz:2018zan,Liu:2019knx} and
a massive dark photon itself being the DM \cite{Nelson:2011sf,Arias:2012az,Graham:2015rva,Nakai:2020cfw}.
We will investigate how to probe this freeze-in DM via the dark photon signals at the CERN SHiP \cite{Alekhin:2015byh,SHiP:2015vad,SHiP:2025ows}, a proposed proton dump experiment which provides access to new freeze-in DM parameter regions not yet explored.

\section{Freeze-in dark matter model via massive dark photon portal}
\subsection{The Model}
We consider a massive dark photon mediated DM model with the following Lagrangian 
\begin{eqnarray}\label{Lag1}
\mathcal{L}_{\rm{dark}}&=&\bar{\chi}\left(i\gamma^{\mu}\partial_{\mu}-m_{\chi}\right)\chi-\frac{1}{4}F'_{\mu\nu}F^{'\mu\nu}-\frac{m^{2}_{A'}}{2}A'_{\mu}A^{'\mu}\nonumber\\
&-&\frac{\epsilon}{2\cos\theta_{W}}F'_{\mu\nu}B^{\mu\nu}+e'J'_{\mu}A^{'\mu}+gJ_{\mu}B^{\mu},
\end{eqnarray}
where fermion $\chi$ is the DM with mass $m_{\chi}$, 
$A'$ is the dark photon with mass $m_{A'}$,
$e(e')$ the (dark) electric charge, 
$B_{\mu}(B_{\mu\nu})$ the SM hypercharge gauge field (strength) with gauge coupling $g$,
$\epsilon$ the small kinetic mixing term,
$J_{\mu}$ the SM hypercharge current, 
and $J'_{\mu}=\bar{\chi}\gamma_{\mu}\chi$ the DM current.

Making a rotation of the fields $A'$ and $B$ to eliminate the kinetic term and subsequent mass mixing in Eq.(\ref{Lag1}), 
one obtains in the broken electroweak phase \cite{Fabbrichesi:2020wbt} 
\begin{eqnarray}\label{Lag2}
\mathcal{L}_{\rm{dark}}&=&\bar{\chi}\left(i\gamma^{\mu}\partial_{\mu}-m_{\chi}\right)\chi-\frac{1}{4}F'_{\mu\nu}F^{'\mu\nu}-\frac{m^{2}_{A'}}{2}A'_{\mu}A^{'\mu}\nonumber\\
&+&e'J'_{\mu}\left(A'^{\mu}+\epsilon \tan\theta_{W}Z^{\mu}\right)-e\epsilon J_{\mu}A'^{\mu}+\mathcal{O}(\epsilon^{2}).\nonumber\\
\end{eqnarray}
The model parameters in Eq.(\ref{Lag2}) are composed of $\{\epsilon, e', m_{\chi}, m_{A'}\}$.

The existing constraints, as seen in Ref.\cite{Fabbrichesi:2020wbt}, allow only $\epsilon<<1$  depending on $m_{A'}$.
With $\epsilon$ being small enough, 
the DM can freeze-in to obtain the observed relic abundance,
depending on the magnitudes of $\epsilon$ and $e'$ as well as the mass parameters $m_{A'}$ and $m_{\chi}$.

\subsection{Sequential freeze-in}
\label{DMrelic}
Now we discuss the DM freeze-in production to address the observed DM relic abundance $\Omega_{\chi}h^{2}=0.12\pm 0.001$ \cite{Planck:2018vyg}.
If $\epsilon>\epsilon_{\rm{th}}\sim 10^{-8}$ \cite{Hambye:2019dwd}, 
where $\epsilon_{\rm{th}}$ is the threshold value required by thermal equilibrium,
the decay of thermalized dark photon into a DM pair is the main DM production process,
leading to $e'\sim 10^{-11}$. In contrast, if $\epsilon<\epsilon_{\rm{th}}$, 
the heavy dark photons are generated from the inverse decay $\psi\bar{\psi}\rightarrow A'$ where $\psi$ denotes electrically charged SM particles,
and subsequently decay to DM particles with $e'\leq\epsilon<\epsilon_{\rm{th}}$, 
which is called as sequential freeze-in DM \cite{Hambye:2019dwd}.

The Boltzmann equations describing the sequential freeze-in process are given by
\begin{eqnarray}
\dot{n}_{A'}+3Hn_{A'}&=&\sum_{\psi}\int d\Pi_{A'}d\Pi_{\psi}d\Pi_{\bar{\psi}}(2\pi)^{4}\delta^{4}(p_{A'}-p_{\psi}-p_{\bar{\psi}})\nonumber\\
&\times&\mid M\mid^{2}_{A'\rightarrow \psi\bar{\psi}}\left(f^{\rm{eq}}_{\psi}f^{\rm{eq}}_{\bar{\psi}}-f_{A'}\right),\label{Boltz1}\\
\dot{n}_{\chi}+3Hn_{\chi}&=&\int d\Pi_{A'}d\Pi_{\chi}d\Pi_{\bar{\chi}}(2\pi)^{4}\delta^{4}(p_{A'}-p_{\chi}-p_{\bar{\chi}})\nonumber\\
&\times&\mid M\mid^{2}_{A'\rightarrow \chi\bar{\chi}} f_{A'}, \label{Boltz2}
\end{eqnarray}
where the sum is over the electrically charged SM particles including leptons and quarks if kinetically allowed, 
$f^{\rm{eq}}$ denotes the Maxwell-Boltzmann distribution function, 
$f_{A'}$ is out-of-equilibrium dark photon distribution function,
$H$ is the Hubble parameter, 
$p_i$ and $\Pi_{i}$ is the momentum and phase-space element of particle $i$, respectively,
and $M$ is decay amplitude.
In Eq.(\ref{Boltz1}), the invisible decay mode being subdominant relative to the visible mode with $e'<\epsilon$ has been neglected; see Sec.\ref{decay} for details.

\begin{figure}
\centering
\includegraphics[width=8cm,height=9cm]{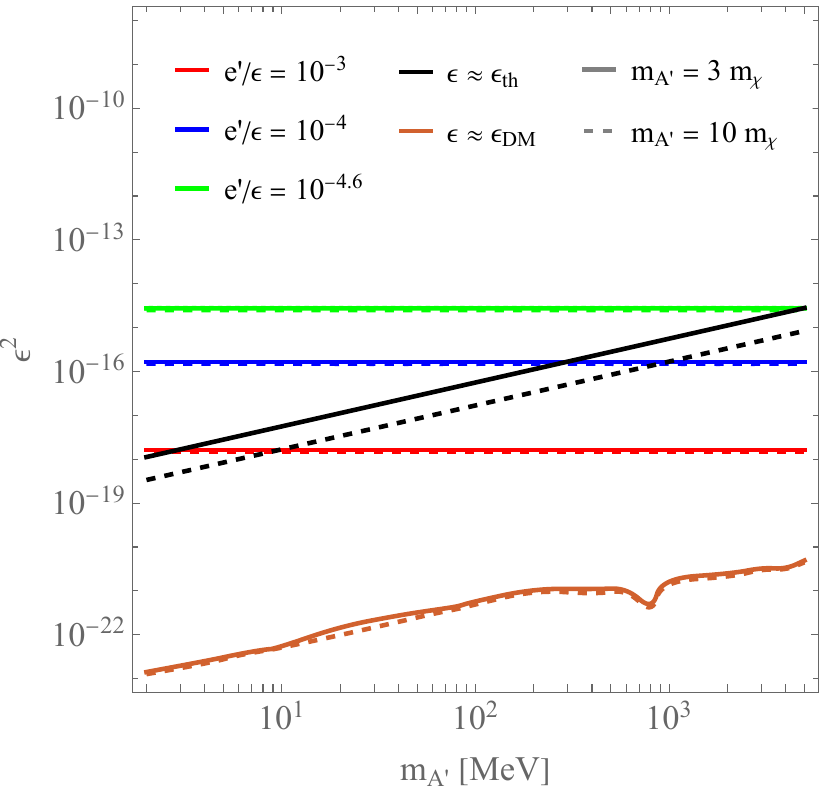}
\centering
\caption{The observed DM relic abundance arising from the sequential freeze-in production projected to the place of $m_{A'}-\epsilon$ for various ratios of $e'/\epsilon=\{10^{-4.6},10^{-4},10^{-3}\}$ and two different mass ratios of $m_{A'}/m_{\chi}=\{3, 10\}$.}
\label{relicdensity}
\end{figure}

To derive the DM relic abundance, it is more convenient to use yields $Y_{A'}$ and $Y_{\chi}$ instead of $n_{A'}$ and $n_\chi$ in Eqs.(\ref{Boltz1}) and (\ref{Boltz2}).\footnote{The $f_{A'}$ term in Eq.(\ref{Boltz1}), often neglected in the literature,
is needed in order to make sure the convergence of the DM yield at $T<< m_{\chi}$.}
We refer the reader to the Appendix for more details. 
Solving Eqs.(\ref{Boltz1}) and (\ref{Boltz2}) numerically,
we show in Fig.\ref{relicdensity} the observed DM relic density projected to the plane of $m_{A'}-\epsilon$ 
for various ratios of $e'/\epsilon=\{10^{-4.6},10^{-4},10^{-3}\}$ and two different mass ratios of $m_{A'}/m_{\chi}=\{3, 10\}$.
Three comments are in order regarding Fig.\ref{relicdensity}:
\begin{itemize}
\item The observed DM relic abundance fixes $e'$ to $\sim 1.3\times 10^{-12}$ as long as $\epsilon\geq \epsilon_{\rm{DM}}$ (in orange),
with $\epsilon_{\rm{DM}}\sim (1-7)\times10^{-11}$ depending on $m_{A'}$.
The nearly fixed value of $e'$, as reflected by the plots with respect to different ratios of $e'/\epsilon$,
is a result of the integral in Eq.(\ref{Y}) being almost a constant as derived from Eq.(\ref{Boltz6}).
\item Second, the value of $\epsilon$ cannot exceed the thermal threshold value $\epsilon_{\rm{th}}$ (in black),
above which the regions are excluded.
\item Third, with the constant integral in Eq.(\ref{Y}), Eq.(\ref{rho}) implies a weak dependence of the DM relic abundance on both $m_{A'}$ and $m_{\chi}$, verified by a change of the mass ratio of $m_{A'}/m_{\chi}$.
\end{itemize}
To summarize, the sequential freeze-in DM parameter space corresponds to $e'\sim 1.3\times 10^{-12}$ and 
$\epsilon_{\rm{DM}}\leq\epsilon<\epsilon_{\rm{th}}$.

Given an effective coupling $\kappa_{\rm{eff}}\sim \epsilon e' \leq 10^{-20}$ between the SM sector and DM,
the DM with the mass range of $m_{\chi}\sim 10^{-2}-10$ GeV is beyond the reaches of any existing direct and indirect detections, which will be simply neglected in the following discussion.

\section{Dark photon signal at the SHiP Experiment}
\label{SHiP}

\subsection{Production}
Similar to the proton bremsstrahlung (Pbrem) process $pp\rightarrow ppA'$, 
dark photons can be generated by the DM bremsstrahlung process of $pp\to \chi \bar{\chi} A^\prime$.
Following Refs.\cite{Blumlein:2013cua,Kriukova:2024wsi}, 
the cross section for this process can be written as
\begin{align}\label{cs}
d\sigma_{pp\to \chi \bar{\chi} A^\prime}&=&\omega_{pA'}(z_1,p_{1\perp}^2)dz_1 dp_{1\perp}^2 d\sigma_{pp\to \chi \bar{\chi}} \nonumber\\
&+&\omega_{\chi A'}(z_2,p_{2\perp}^2)dz_2 dp_{2\perp}^2 d\sigma_{pp\to \chi \bar{\chi}},
\end{align}
where $\omega_{xA'}$ denotes the probability for an $A'$ to be radiated from particle $x$, commonly referred to as the splitting function, 
the dark photon momentum perpendicular and parallel to the beam direction is denoted by $p_\perp$ and $p_\parallel$ respectively, 
and the parameter $z$ defines the fraction of the $x$'s longitudinal momentum transferred to the dark photon.

\begin{figure}
\centering
\includegraphics[width=8cm,height=8cm]{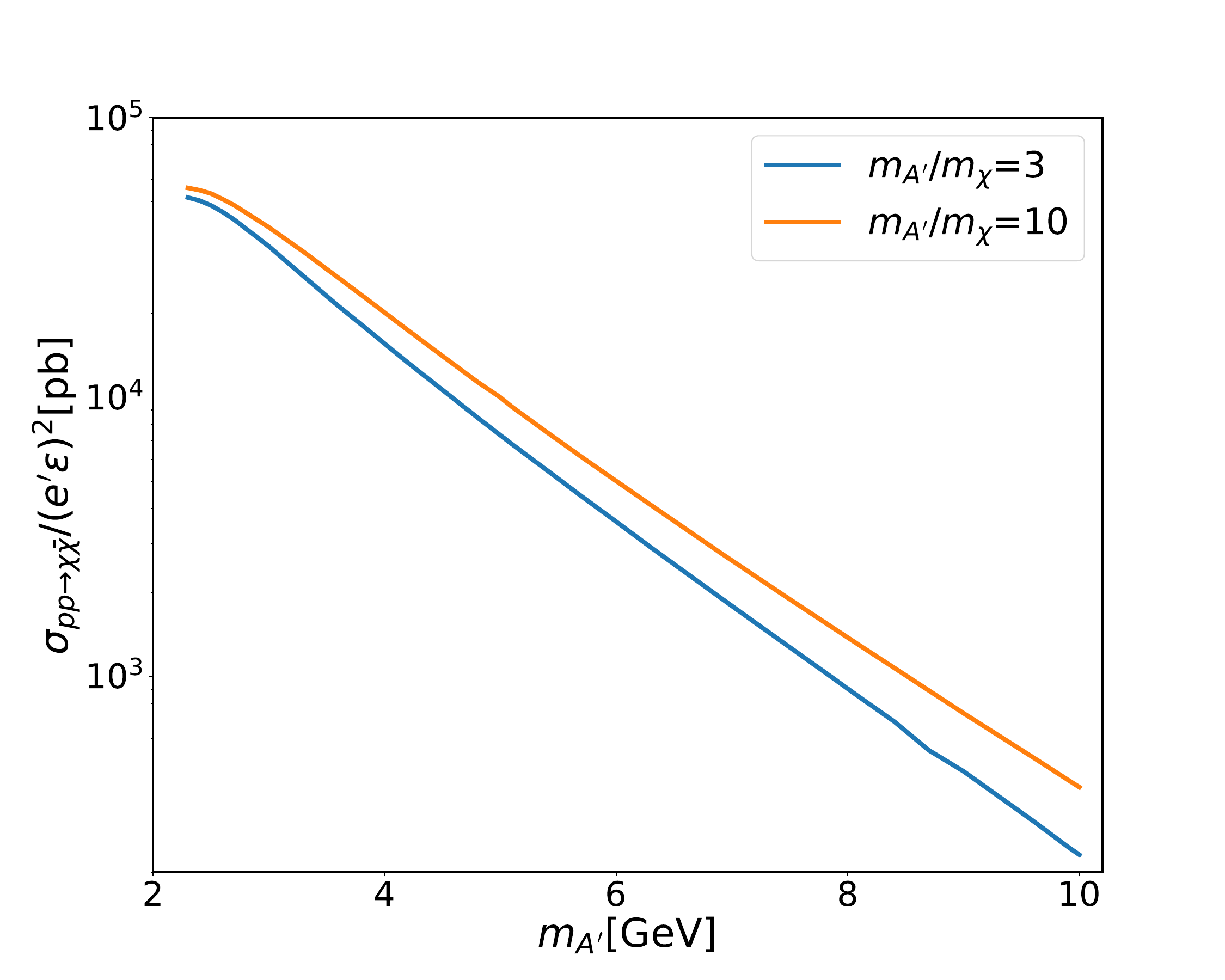}
\centering
\caption{ The values of $\sigma_{pp\to \chi \chi}/(e'\epsilon)^{2}$ in Eq.(\ref{cs}) for two different mass ratios of $m_{A'}/m_{\chi}=\{3,10\}$.}
\label{sigma}
\end{figure}

Put aside the $\omega_{xA'}$ factor,
Fig.\ref{sigma} shows the values of $\sigma_{pp \to \chi \bar{\chi}}$ in Eq.(\ref{cs}), which scales as $\sim \epsilon^{2}e'^{2}$, 
with one $p$ having an energy of $400$ GeV while the other $p$ at rest.
For  $m_{\chi}$ above $\sim 2$ GeV, we calculate $\sigma_{pp \to \chi \bar{\chi}}$ using MadGraph5\_aMC@NLO \cite{Alwall:2014hca}.
For  $m_{\chi}$ below $\sim 2$ GeV,  $\sigma_{pp \to \chi \bar{\chi}}$ further increases as $m_{A'}$ decreases. 
However, taking into account the suppression due to $e'$ and $\epsilon$, 
$\sigma_{pp \to \chi\bar{\chi}}$ is at least several orders of magnitude smaller than $\sigma^{inel}_{\rm{SHiP}}\approx 10.7$ mb \cite{SHiP:2015vad},
the reference cross section related to the SM meson and Pbrem process for the dark photon.
Because of the severe suppression, we ignore the DM bremsstrahlung contribution to the dark photon production at the SHiP.

\subsection{Decay}
\label{decay}
The heavy dark photon decays into lepton pair $\ell^{+}\ell^{-}$ with $\ell=\{e,\mu,\tau\}$, hadronic states contributing to electrically charged particles, and DM pair $\chi \bar{\chi}$.

For the lepton decay channels, the partial decay width is given by \cite{Blumlein:2013cua}
\begin{eqnarray}\label{Wl}
\Gamma\left(A' \to \ell^{+}\ell^{-}\right)=\frac{1}{3}\alpha\epsilon^2 m_{A'} \sqrt{1-\frac{4m_{\ell}^2}{m_{A'}^2}} \left( 1+\frac{2m_{\ell}^2}{m_{A'}^2} \right),\nonumber\\
\end{eqnarray}
where $\alpha$ is the fine structure constant and $m_{\ell}$ is the lepton mass.

Regarding the hadron decay channels, the partial decay width reads as \cite{Bjorken:2009mm}
\begin{eqnarray}\label{Wh}
\Gamma(A'\to \rm{hadrons})=\Gamma(A' \to \mu^{+} \mu^{-})R,
\end{eqnarray}
where
\begin{eqnarray}\label{R}
R=\frac{\sigma(e^+ e^- \to \rm{hadrons})}{\sigma(e^+ e^- \to \mu^+ \mu^-)},
\end{eqnarray}
is an energy-dependent ratio \cite{ParticleDataGroup:2014cgo}.

For the DM decay channels, the partial decay width is
\begin{eqnarray}\label{Wd}
\Gamma(A'\to \chi \bar{\chi})=\frac{1}{3}\alpha_{D}m_{A'}\sqrt{1-\frac{4m_{\chi}^2}{m_{A'}^2}} \left( 1+\frac{2m_{\chi}^2}{m_{A'}^2} \right),
\end{eqnarray}
for $m_{A'}>2m_{\chi}$, with $\alpha_{D}={e^{\prime}}^2/ 4\pi$.
With $e'/\epsilon\geq 1$, the invisible decay mode dominates over the visible decay modes.
In contrast, given $e'/\epsilon\leq 0.1$, the invisible decay mode can be safely neglected.

\subsection{Signal sensitivities}
\label{signal}
The number of events detected by the SHiP experiment is given by 
\begin{eqnarray}\label{Events}
\mathcal{N}_{A'}=\sum_{i}\sigma_{i}\times \mathcal{L}_{\rm{SHiP}}\times \rm{Br}(A'\rightarrow \rm{ch}+\rm{ch})\times \mathcal{P}_{\rm{v},i}\times \mathcal{P}_{\rm{r},i},\nonumber\\
\end{eqnarray}
where the summation is over the meson, Pbrem and Drell-Yan processes,
$\mathcal{L}_{\rm{SHiP}}=N/\sigma^{inel}_{\rm{SHiP}}$ with $N$ the number of protons on the target, 
$\rm{Br}(A'\rightarrow \rm{ch}+\rm{ch})$ is the branching ratio excluding the neutral hadron final states,
and finally $\mathcal{P}_{\rm{v}}$ and $\mathcal{P}_{\rm{r}}$ is the detector acceptance and efficiency \cite{SHiP:2020vbd} respectively.

\begin{figure*}[t]
\includegraphics[width=16cm,height=13cm]{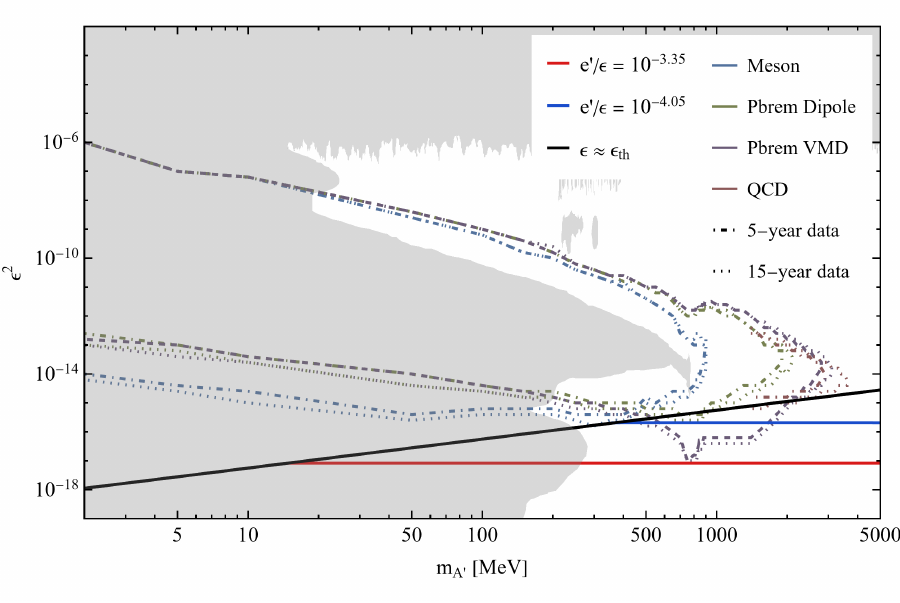}
\caption{The sequential freeze-in DM parameter space corresponding to $\epsilon_{\rm{DM}}\leq\epsilon<\epsilon_{\rm{th}} (\rm{in~black})$ compared to 
the 90\% CL limits of the meson, Pbrem (dipole $\&$ VMD) and QCD production processes for the dark photon with $N=\{2, 6\}\times 10^{20}$,
and the existing bounds (in shaded gray region). See text for details.}
\label{limits}
\end{figure*}

Consider that the SHiP limits on $\epsilon$ are sensitive to the value of $N$.
We extend the analysis of the 90\% confidence level (CL) limits with respect to $N=2\times 10^{20}$ (5-year operation) in Ref.\cite{SHiP:2020vbd} to $N=6\times 10^{20}$ (15-year operation).
To do so we use the values of cross sections $\sigma_{i}$ and the distributions of $\mathcal{P}_{\rm{v}}$ and $\mathcal{P}_{\rm{r}}$ in the two-dimensional plane of $m_{A'}-\epsilon$ therein to extract the new samples satisfying the number of events $\mathcal{N}_{A'}\geq 2.3$.

Fig.\ref{limits} presents the 90\% CL limits of three different production processes with $N=\{2, 6\}\times 10^{20}$ on the sequential freeze-in DM 
parameter space described by the contours of $e'/\epsilon=\{10^{-4.05}, 10^{-3.35}\}$ with $m_{A'}=3m_{\chi}$,
which are compared to the existing bounds (shaded region in gray) and the thermal condition on $\epsilon_{\rm{th}}$ (in black).
Here, only the $\epsilon_{\rm{th}}$ curve is sensitive to the mass ratio of $m_{A'}/m_{\chi}$ as illustrated by Fig.\ref{relicdensity},
the Pbrem process is divided into the dipole dominance and vector meson dominance (VMD) case respectively,
and the existing bounds arise from NA64 \cite{NA64:2019auh}, LHCb \cite{LHCb:2019vmc},  BaBar \cite{BaBar:2014zli},
E137 \cite{Batell:2014mga} , CHARM \cite{CHARM:1985anb}, NuCal \cite{Gninenko:2012eq} etc; 
see Refs.\cite{Fabbrichesi:2020wbt, Beacham:2019nyx} for a complete list of the references.

\section{Discussion}
Massive dark photon mediated DM, 
being active within the region of $\epsilon< \epsilon_{\rm{th}}$, 
has been frequently mentioned in the literature.
However, the precise information on such DM is missing. 
This study has filled the gap by pointing out the concrete parameter space of sequential freeze-in DM which 
corresponds to $e'\sim 1.3\times 10^{-12}$ and $\epsilon_{\rm{DM}}\leq\epsilon<\epsilon_{\rm{th}}$, 
with $\epsilon_{DM}\sim 10^{-11}$ and $\epsilon_{\rm{th}}\sim 10^{-8}-10^{-7.5}$ for the dark photon mass $m_{A'}\sim 10^{-2}-10$ GeV, as seen in Fig.\ref{relicdensity}.

Besides, we have shown that a large portion  of the $\epsilon$ region allowed is in the reach of the proposed CERN SHiP experiment. 
Concretely speaking, the 5(15)-year observations of the Pbrem production process for the dark photon,  
assuming VMD (dipole dominance), 
exclude $\epsilon\geq 10^{-8.5} (10^{-7.9})$ at 90\% CL, as shown by the red (blue) plot in Fig.\ref{limits}.
Only a narrow region of $\epsilon$ close to $\sim 10^{-11}$ is left for alternative tests.

$Acknowledgements$. The authors thank A. M. Magnan for discussion about the data of SHiP detector acceptance and efficiency.

\pagebreak
\widetext
\begin{center}
\textbf{\large Appendix}
\end{center}
\setcounter{equation}{0}
\setcounter{figure}{0}
\setcounter{table}{0}
\setcounter{section}{0}
\makeatletter
\renewcommand{\theequation}{S\arabic{equation}}
\renewcommand{\thefigure}{S\arabic{figure}}
\renewcommand{\bibnumfmt}[1]{[#1]}
\renewcommand{\citenumfont}[1]{#1}

\hspace{5mm}

This appendix provides the details of calculating the DM relic abundance due to the sequential freeze-in process of $\psi\bar{\psi}\rightarrow A'\rightarrow \chi\bar{\chi}$. 

Using the Maxwell-Boltzmann distribution function 
\begin{eqnarray}
\label{thermal}
f^{\rm{eq}}\approx e^{-E/T},
\end{eqnarray}
where $T$ is the temperature of the SM thermal bath,
Eq.(\ref{Boltz1}) can be rewritten as 
\begin{eqnarray}
\dot{n}_{A'}+3Hn_{A'}=\frac{g_{A'}}{2\pi^{2}}m^{2}_{A'}TK_{1}(m_{A'}/T)\sum_{\psi}\Gamma_{A'\rightarrow \psi\bar{\psi}}-g_{A'} \left(\int d\Pi_{A'}\frac{f_{A'}}{\gamma_{A'}}\right)\sum_{\psi}\Gamma_{A'\rightarrow \psi\bar{\psi}}, \label{Boltz3}
\end{eqnarray}
where  $g_{A'}=3$ is the number of degrees of freedom, 
$K_{1}$ is the first modified Bessel function of the 2rd kind,
the summation is over electrically charged leptons and hadrons, 
$\gamma_{A'}=E_{A'}/m_{A'}$ is the $\gamma$-factor,
and the invisible decay mode being sub-dominant compared to the visible modes has been neglected. 
According to the definition of dark photon number density 
\begin{eqnarray}
\label{nA}
n_{A'}=\frac{g_{A'}}{(2\pi)^{3}}\int d^{3}\mathbf{p}_{A'} f_{A'},
\end{eqnarray}
Eq.(\ref{Boltz3}) can be rewritten as
\begin{eqnarray}
\dot{n}_{A'}+3Hn_{A'}=\frac{3}{2\pi^{2}}m^{2}_{A'}TK_{1}(m_{A'}/T)\sum_{\psi}\Gamma_{A'\rightarrow \psi\bar{\psi}}
-n_{A'}\sum_{\psi}\Gamma_{A'\rightarrow \psi\bar{\psi}}. \label{Boltz4}
\end{eqnarray}

Moreover, plugging Eq.(\ref{nA}) into Eq.(\ref{Boltz2}) gives
\begin{eqnarray}
\dot{n}_{\chi}+3Hn_{\chi}\approx \left(\frac{m_{A'}}{m_{\chi}}\right)n_{A'}\Gamma_{A'\rightarrow \chi\bar{\chi}}.\label{Boltz5}
\end{eqnarray}

Transforming the number densities $n_{A'}$ ($n_{\chi}$) to yields $Y_{A'}$ ($Y_{\chi}$) and the time variable to $T$ with $dT/dt\approx -HT$ in the radiation dominated epoch,  and using $S=2\pi^{2}g_{*}^{S}T^{3}/45$ and $H\approx 1.66\sqrt{g_{*}^{\rho}}T^{2}/M_{P}$ with $g_{*}^{S}$ and $g_{*}^{\rho}$ the number of degrees of freedom in the entropy $S$ and the energy density $\rho$ respectively, 
one obtains the final forms of the Boltzmann equations:
\begin{eqnarray}
\frac{dY_{A'}}{dx}&\approx&\mathcal{C}_{1A'} x^{3}K_{1}(x)\sum_{\psi}\Gamma_{A'\rightarrow \psi\bar{\psi}}-\mathcal{C}_{2A'} x Y_{A'}(x)\sum_{\psi}\Gamma_{A'\rightarrow \psi\bar{\psi}}, \label{Boltz6}\\
\frac{dY_{\chi}}{dx}&\approx& \mathcal{C}_{\chi} xY_{A'}(x)\Gamma_{A'\rightarrow \chi\bar{\chi}}, \label{Boltz7}
\end{eqnarray}
where $x=m_{A'}/T$ is a dimensionless variable, and the coefficients 
\begin{eqnarray}
\mathcal{C}_{1A'}&=&\frac{45 g_{A'}}{1.66(2\pi^2)^2}\frac{M_{P}}{\sqrt{g_{*}^{\rho}} g_{*}^{S} m_{A'}^2}\label{coef1},\\
\mathcal{C}_{2A'}&=&\frac{M_{P}}{1.66 \sqrt{g_{*}^{\rho}}}\frac{1}{m_{A'}^2}\label{coef2},\\
\mathcal{C}_{\chi} &=& \frac{M_{P}}{1.66 \sqrt{g_{*}^{\rho}}} \frac{1}{m_{\chi}m_{A'}}. \label{coef3}
\end{eqnarray}

Solving Eqs.(\ref{Boltz6}) and (\ref{Boltz7}) numerically,
one gets the present-day value of DM yield 
\begin{eqnarray}
\label{Y}
Y_{\chi}(\infty)\approx\mathcal{C}_{\chi}\Gamma_{A'\rightarrow \chi\bar{\chi}} \int^{\infty}_{0} xY_{A'}(x)dx,
\end{eqnarray}
resulting in the DM energy density
\begin{eqnarray}
\label{rho}
\rho_{\chi}=Y_{\chi}(\infty)S_{0}m_{\chi},
\end{eqnarray}
where $S_0$ is the present value of entropy.

\end{document}